\documentclass[journal,twocolumn,10pt,twoside]{IEEEtran}

\usepackage{cite}

\ifCLASSINFOpdf
\else
\fi

\usepackage{url}
\usepackage{amsmath}
\usepackage{amssymb}
\usepackage{cite}
\usepackage{graphicx}
\usepackage{algorithm}
\usepackage{algpseudocode}
\usepackage{subfigure}
\usepackage{multirow} 
\usepackage{longtable}
\usepackage{threeparttable}
\usepackage{caption3}
\usepackage{array}
\usepackage{cases}
\usepackage{color}
\usepackage[T1]{fontenc}
\usepackage[utf8]{inputenc}
\usepackage{authblk}
\usepackage{subfigmat}
\usepackage{caption}
\usepackage{multirow}
\usepackage{float}
\usepackage{multirow, makecell}
\usepackage{booktabs}
\usepackage{epstopdf}

\begin{document}

\setlength{\skip\footins}{2mm}
\addtolength{\textfloatsep}{-7mm}
\addtolength{\floatsep}{-5mm}

\title{Range Expansion for Wireless Power Transfer using Joint Beamforming and Waveform Architecture:\\ An Experimental Study in Indoor Environment}
%
%
\author{Junghoon~Kim~and~Bruno~Clerckx
\thanks{J. Kim and B. Clerckx are with the Department
of Electrical and Electronic Engineering, Imperial College London, London,
SW7 2AZ, UK, e-mail: (junghoon.kim15,b.clerckx@imperial.ac.uk).

This work has been partially supported by the Engineering and Physical Sciences Research Council (EPSRC) of U.K. under Grant EP/P003885/1 and EP/R511547/1.
}
}

\maketitle

\begin{abstract}
Far-field Wireless Power Transfer (WPT) has emerged as a potential power source for the Internet of Things (IoT) and Wireless Sensor Network (WSN).
The expansion of the power transfer range is one of the key challenges to make the technology viable.
In this paper, we experimentally study a channel-adaptive joint beamforming and waveform architecture to expand the power transfer range.
WPT experiments have been conducted in a variety of wireless channels at various distances in a realistic indoor environment. 
The measurement data have been fitted using a simple analytical model to analyze the output DC power and achievable range improvement depending on the signal design schemes and the number of tones and antennas. 
The model shows a clear relationship between signal design versus output DC power and achievable range, and highlight the significant benefit of the proposed architecture to expand the power transfer range.
\end{abstract}


\IEEEpeerreviewmaketitle

\vspace{-0.4 cm}

\section{Introduction}

\IEEEPARstart{F}{ar} field Wireless Power Transfer (WPT) using radio-frequency (RF) signals is attracting attention as a viable power source for Internet of Things (IoT) and wireless sensor network \cite{Visser2013}, and the enabler of 1G mobile power network \cite{Clerckx2018b}.
The most significant benefit of RF WPT is that energy can be transferred over longer distances in line-of-sight (LoS) and non-line-of-sight (NLoS) deployments and receiver can be implemented in a smaller form factor compared to other technologies.
The crucial challenge of RF WPT is to increase the harvested DC power at the receiver without increasing the transmit power and to expand the power transfer range while maintaining a power level sufficient to run applications.
To that end, there have been a lot of efforts in the RF community to design high-efficiency rectennas \cite{Valenta2014}, and in the communication community to design optimal transmission signals which exploit multi-antenna beamforming and waveform, channel state information (CSI), and rectifier characteristics \cite{Zeng2017}. 
\par 
Expansion of wireless power transfer range using waveform design has been studied and implemented in RFID applications \cite{Trotter2010, Boaventura2013a}.
Received DC power improvement by multi-tone power optimized waveforms (POW) was confirmed by simulations and RFID read range experiments.
The POW signals used multi-tone strategy but it was a pre-designed waveform and not adaptive to the wireless channel. 
A critical step forward in signal design was made in \cite{Clerckx2016}, where a nonlinear rectifier model was proposed and the optimal channel-adaptive joint waveform and beamforming design for WPT was derived. 
Such optimized signals were then shown theoretically to extend the range of WPT \cite{Huang2017}.
Waveform design techniques that can be used in a real-time system by reducing the complexity have also been proposed in \cite{Clerckx2017}.
There also have been studies on WPT system implementation in a realistic environment. 
The first prototype of a WPT system with CSI acquisition and the channel-adaptive waveform was implemented in \cite{Kim2017}.
Closed-loop WPT systems with adaptive multi-antenna beamforming in realistic channels have been implemented in \cite{Choi2018, Yedavalli2017}. 
More recently, WPT performance benefits of joint channel-adaptive beamforming and waveform have been experimentally evaluated and confirmed in realistic wireless channels \cite{Kim2020}.
\par
In this paper, instead of focusing on showing the benefits of channel-adaptive signal designs to boost the output DC power performance as in \cite{Kim2017,Kim2020, Choi2018,Yedavalli2017}, we investigate how joint waveform and beamforming can be used to expand the WPT range for a fixed power delivery.  
Expanding the range of WPT is crucial to enable a wide deployment of low power wireless-powered applications.
The main contributions of the paper are summarized as follows.
\par
\textit{First}, we implement an improved prototype of a WPT system relying on CSI acquisition and joint waveform and beamforming design that can support a larger number of transmit antennas and over-the-air CSI feedback, and established a WPT testbed in a real indoor office environment.
Downlink CSI is estimated using pilot signal and reported back to the transmitter by uplink over-the-air feedback\footnote{Prototype in \cite{Kim2017,Kim2020} were limited to a smaller number of antennas and the CSI feedback was based on cable transmission, instead of over-the-air.}. 
We experimentally evaluate the output DC power performance and power transfer range of each signal design scheme under various wireless channel conditions and the same power constraint. 
\par
\textit{Second}, the measurement results in a real-world environment have been collected and used to build simple analytical models using curve-fitting. 
Measurement and fitting results enable us to analyse and predict how the distance as well as the transmit signal design properties, such as the number of frequency components and the number of transmit antennas, affect the WPT output DC performance.
Based on the analysis, we discuss how the WPT range can be expanded using proper signal design strategies. 
Approximately, a power transfer range gain of 15\% and 60\% can be obtained when the number of tones or antennas are doubled, respectively. 
Those gains are cumulative using joint beamforming and waveform signal strategies; for example, using a channel-adaptive 4-antenna and 8-tone signal leads to approximately 400\% of range gain over continuous-wave (CW). 
This is the first paper that experimentally demonstrates the significant range expansion offered by joint waveform and beamforming design for WPT. 
\par
\textit{Organization}: Section \ref{signal} introduces the system model and Section \ref{testbed} presents the prototype and testbed implementation. Section \ref{analysis} provides all experimental results and an analytical model based on curve-fitting. Section \ref{conclusion} concludes the work.

\textit{Notations}: Bold letters stand for vectors or matrices whereas a symbol not in bold font represents a scalar. $\left|.\right|$ and $\left\| . \right\|$ refer to the absolute value of a scalar and the 2-norm of a vector. $\mathbb{E}\{ .\}$ refers to the averaging/expectation operator.

\vspace{-0.3 cm}
\section{WPT System and Signal Models} \label{signal}

We consider a Multiple Input-Single Output (MISO) point-to-point WPT system using various channel adaptive and non-adaptive waveforms for WPT introduced in \cite{Clerckx2016}. 
In this section, we briefly describe the WPT system model, including the nonlinear rectifier model, as well as channel adaptive WPT signal design schemes and their performance scaling laws. 

\vspace{-0.3 cm}
\subsection{MISO WPT System model}
The transmitter is equipped with $M$ antennas in the MISO WPT system, and the transmission signals from each antenna use $N$ frequency components. 
The vector-wise multi-tone transmission signal comprises $M$ signals from different antennas can be written as

\begin{equation}
 \mathbf{x}(t) = \Re\Bigg\{ \sum_{n=0}^{N-1}\mathbf{w}_{n}e^{j2\pi f_{n}t}\Bigg\}
\end{equation}
where $\mathbf{w}_{n}{=}\left[ \omega_{n,1}  \cdots  \omega_{n,M} \right]^{T}$ with $\omega_{n,m}{=}s_{n,m}e^{j\phi_{n,m}}$, $s_{n,m}$ and $\phi_{n,m}$ refer to the amplitude and phase of the single-tone signal on frequency $f_{n}$ and transmit antenna $m$.
Frequency components are evenly spaced, i.e., $f_{n}{=} f_{0}+n\Delta_{f}$.
The average transmit power constraint is given by $\sum_{m=1}^{M}\mathbb{E}\{ \left|x_{m}\right|^{2}\} \le P$, where $x_{m}$ is the transmit signal at antenna $m$. 

\par
The signal from antenna $m$ is transmitted to the receiver through the multipath wireless channel denoted by $ \Lambda ^{-1/2}h_{n,m} $ where $ \Lambda$ is the path loss and $h_{n,m}$ is the complex fading coefficient for frequency component $n$ (we omit the time dependency based on the assumption that the channel changes slowly).
The total received signal is the sum of the received signals from different transmit antennas, namely

\begin{align}
\begin{split}
 y(t) &= \Re\Bigg\{ \sum_{n=0}^{N-1}\Lambda ^{-1/2}\mathbf{h}_{n}\mathbf{w}_{n}e^{j2\pi f_{n}t}\Bigg\}.
\end{split}
\end{align}
where $\mathbf{h}_{n} {=} \left[ h_{n,1}  \cdots h_{n,M}  \right]$ which is a channel vector for the frequency component $n$ for all antennas.
Each fading coefficient $h_{n,m}$ consists of amplitude and phase $A_{n,m}$ and $\bar{\psi}_{n,m}$ such that $h_{n,m} {=} A_{n,m}e^{j\bar{\psi}_{n,m}}$.

\vspace{-0.2 cm}
\subsection{WPT Waveform Designs and Non-linear Rectifier Model}
We consider several waveform design schemes based on channel non-adaptive waveform and channel-adaptive waveform.
Channel non-adaptive CW is the simplest signal design that assigns all available power to only one frequency and antenna component, so only $\omega_{1,1}$ has an amplitude value of $\sqrt{2P}$ and the remaining components have zero value. 
We use scaled matched filter (SMF) signal design method for generating channel-adaptive WPT waveform. 
The SMF is a low-complexity multi-tone waveform design method proposed in \cite{Clerckx2017,Kim2020}, which aligns phase and allocates power to each frequency and spatial component proportionally to the channel magnitude.
This method applies two WPT signal design strategies such as waveform and beamforming for frequency domain and spatial domain, respectively. 
The SMF has a scaling factor $\beta$, whose choice results from a compromise between exploiting the rectifier nonlinearity and the channel frequency selectivity.  
When the number of tones is fixed to one, the signal is the same as single-tone and multi-antenna beamforming signal with maximal ratio transmission (MRT).  

\par
The received RF signal is converted to DC power through a single-diode rectifier. 
The output DC power can be expressed as a function of the input RF signal $y(t)$ using the simple and tractable nonlinear rectifier model introduced in \cite{Clerckx2016}.
The nonlinear model was confirmed to be more accurate than the classical linear model in various previous studies \cite{Clerckx2016,Clerckx2017,Boaventura2013,Kim2020}.
The output DC power performance at the receiver according to the transmission signal design and wireless channel conditions can be evaluated using the nonlinear rectifier model.
Following \cite{Clerckx2016}, the scaling law of the channel-adaptive signal with large $N$ and $M$ and CW for both frequency-flat (FF) and -selective (FS) channels is given respectively as
\vspace{-0.2 cm}
\begin{align}
& z_{\mathrm{DC,ca}}  &=&\quad   k_{2}R_\mathrm{ant}\Lambda ^{-1}PM + k_{4}R_\mathrm{ant}^{2}\Lambda ^{-2}P^{2}NM^{2} \\
& z_{\mathrm{DC,cw}} &=&\quad  k_{2}R_\mathrm{ant}\Lambda ^{-1}P + 3k_{4}R_\mathrm{ant}^{2}\Lambda ^{-2}P^{2}
\end{align}
where $\Lambda$ is the path loss, $P$ is the power constraint at the transmitter, $R_\mathrm{ant}$ is the receive antenna impedance, and $k_{2, 4}$ are constants from the non-linear rectifier model.
Readers are referred to \cite{Clerckx2016} for more details on the scaling laws. 
The scaling law shows WPT performance of the channel-adaptive signal can theoretically be increased proportionally to the number of transmit antennas $M$ and the number of frequencies $N$ for a given distance $d$. 
The dependency on the distance $d$ between transmitter and receiver is reflected in the path loss $\Lambda$.
In the following sections, we measure the performance in a real wireless environment with various distances, tones and antennas and analyze the effect of each factor $d$, $N$, $M$ on the actual performance. 
We confirm experimentally the theoretical behavior and the fact that the increase in $N$ and $M$ with channel-adaptive signals plays a big role in boosting the output DC power and the power transfer range.

\vspace{-0.3 cm}
\section{WPT Testbed System and Experiment Setup.} \label{testbed}

\begin{figure}
	\centering
	\includegraphics[width=0.4\textwidth]{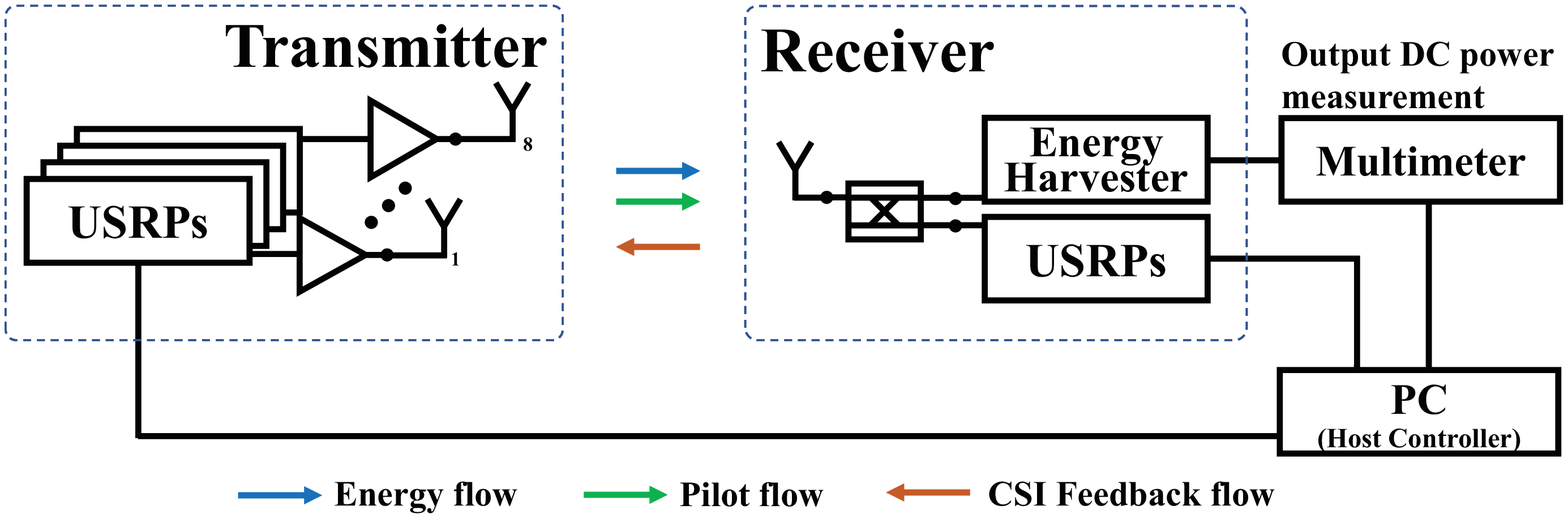}
	\caption{WPT testbed system configuration.}
	\label{diagram}
\end{figure}

We implemented a point-to-point WPT testbed system which supports up to eight transmit antennas and CSI feedback. 
The goal is to leverage and improve the prototype reported in \cite{Kim2020} to analyze how multi-tone, multi-antenna and channel adaptive waveforms can expand the WPT range. 
The WPT system operates at the 2.4GHz ISM band, the pilot signal for channel estimation and the channel adaptive WPT waveforms are transmitted on the downlink (transmitter to receiver) channel, and the estimated CSI is transferred on the uplink (receiver to transmitter) channel.
Some signal design methods presented in the previous section are implemented.
The bandwidth for multi-tone signals is fixed to 10 MHz in this experiment, and the $N$ frequency components are evenly spaced within that bandwidth.
A brief configuration of the WPT testbed system is illustrated in Fig.\ref{diagram}.
\par
\begin{figure}
	\centering
	\subfigure[]{\includegraphics[width=0.17\textwidth]{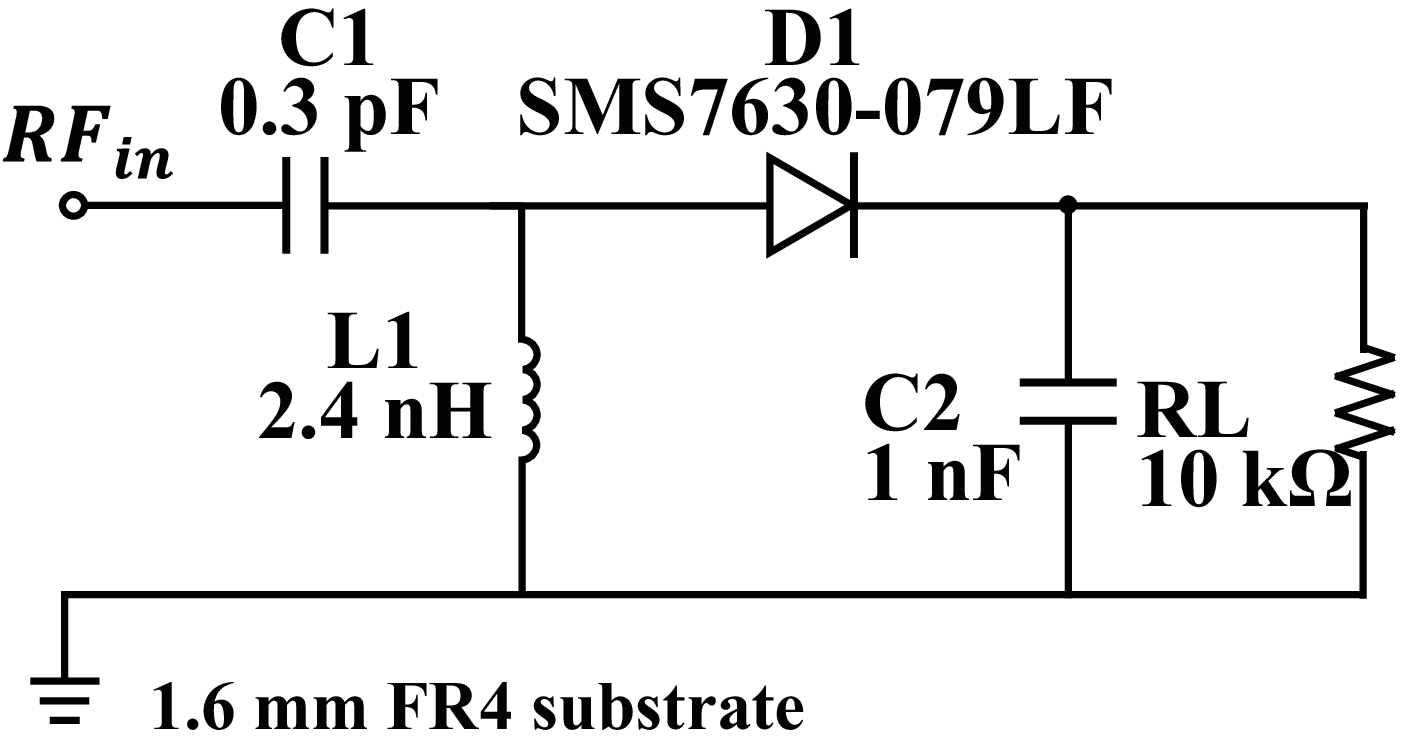}}
	\subfigure[]{\raisebox{2mm}{\includegraphics[width=0.18\textwidth]{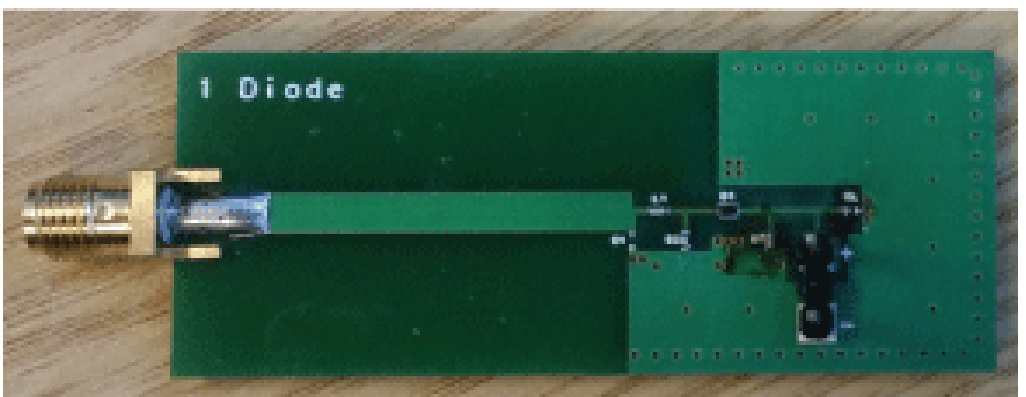}}}
	\caption{Single-diod rectifier design (a) schematic, (b) fabrication.}
	\label{rec}
\end{figure}

The transmitter is implemented using USRPs (NI USRP-2942), supports up to eight transmit antennas, can generate the WPT signal (adaptive to the CSI) as well as a predefined pilot signal, and simultaneously receive CSI feedback from the receiver.
External power amplifiers (ZHL-16W-43+) are applied between the transmit antennas and the USRPs to enable WPT transfer distance around 5 m in an indoor room environment while complying with the maximum EIRP of 36dBm specified in FCC Title 47, Part 15 regulations \cite{FCC}.
The RF signal input to the receiving antenna is distributed to the two different functional blocks via a power splitter\footnote{The received RF power at the energy harvester can be increased by 3dB using RF switch which distributes entire received RF signals to energy harvester when WPT waveforms are input compared to using a power splitter. However, we used power splitter to reduce implementation complexity and to simultaneously monitor received RF power at the USRP. The main objective of this paper is WPT performance comparison between different waveform design schemes, number of frequencies and antennas, so using a power splitter does not affect the observations.}. Those blocks are the energy harvester for converting RF signals to DC power and USRP for channel estimation and feedback. 
We have used the same single-diode rectifier as the energy harvester used in \cite{Kim2020}, reproduced in Fig. \ref{rec}, and the output DC voltage was measured using a multimeter (Keysight 34465A).
The energy harvester simply consists of a matching network, rectifying diode (Skyworks SMS7630) and low pass filter circuits.
The values of the circuit components were chosen to fit the 2.45GHz operation frequency, and the entire circuit is fabricated using 1.6mm thick FR-4 substrate and lumped elements.

\par
The testbed system is operated by time-frames, and a frame is designed to be one second long, such that CSI acquisition and WPT waveform design are performed every one second. 
The frame length was chosen because our experiment is carried out under static conditions where the wireless channel does not change rapidly.
The CSI estimation is carried out at the receiver side USRP using the least-square method and quantized by 16 bits, then the information is modulated using OFDM and transferred over-the-air on the uplink to the transmitter\footnote{CSI estimation and feedback is a power consuming process. We do not consider the power consumption for CSI acquisition in this paper, and the design and implementation of an efficient CSI acquisition system with reduced power consumption remains an interesting future work.}.
The functional block to receive and demodulate the feedback signal is implemented at the transmitter side USRP.
Approximately 80ms is taken to acquire CSI at the transmitter out of one-second frame length. 
The operating sequence of the testbed system is implemented by LabView, and the entire system is controlled by a host PC.
\begin{figure}
	\centering
	\subfigure[]{\includegraphics[width=0.28\textwidth]{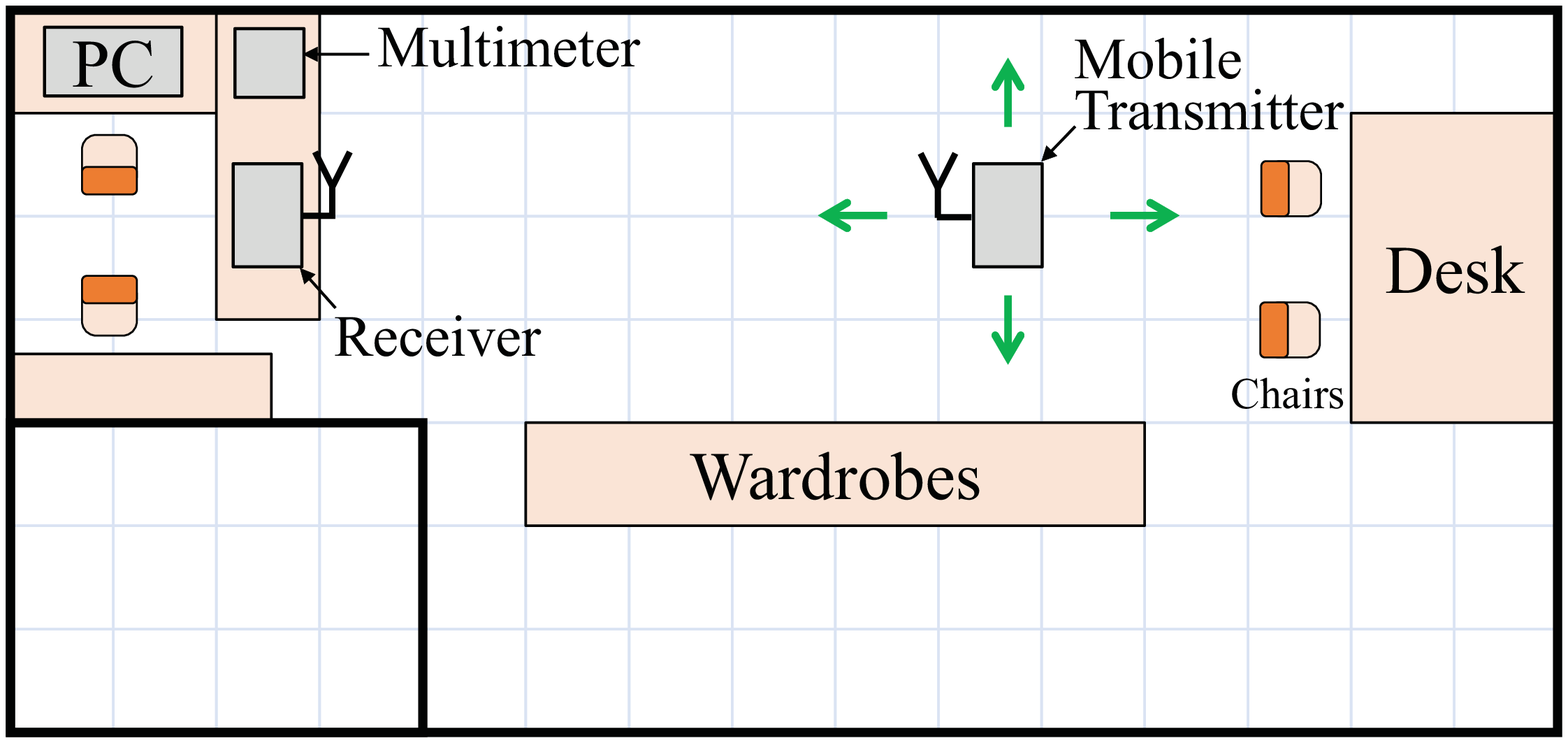}}
	\subfigure[]{\includegraphics[width=0.10\textwidth]{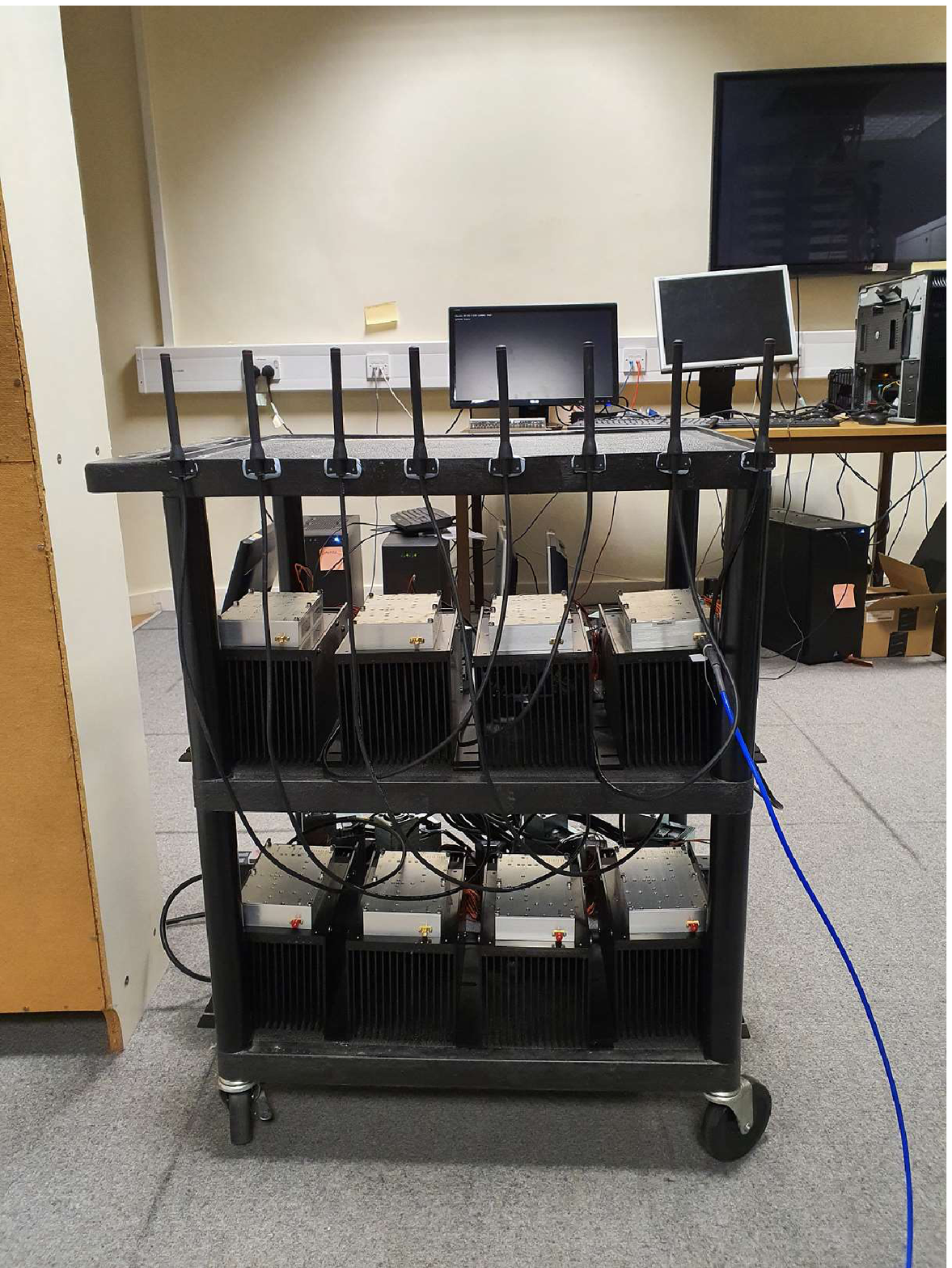}}
	\caption{Installation of WPT testbed system (a) Layout of the indoor laboratory and locations of receiver and transmitter (b) Mobile transmit antennas trolley.}
	\label{layout}
\end{figure}
The testbed system was installed in an indoor laboratory of 7m$\times$5m with common facilities such as a desk, chair, PCs and wardrobes that cause multipath fading in the wireless channel. 
The location of the receiver is fixed but the transmitting antenna is installed on the mobile trolley to be located anywhere in the laboratory.
A brief layout of the indoor laboratory and photo of the transmit antennas are shown in Fig. \ref{layout}.

\begin{figure*}[t] 

  \begin{minipage}[b]{0.32\linewidth}
	\centering
	\includegraphics[width=0.91\textwidth]{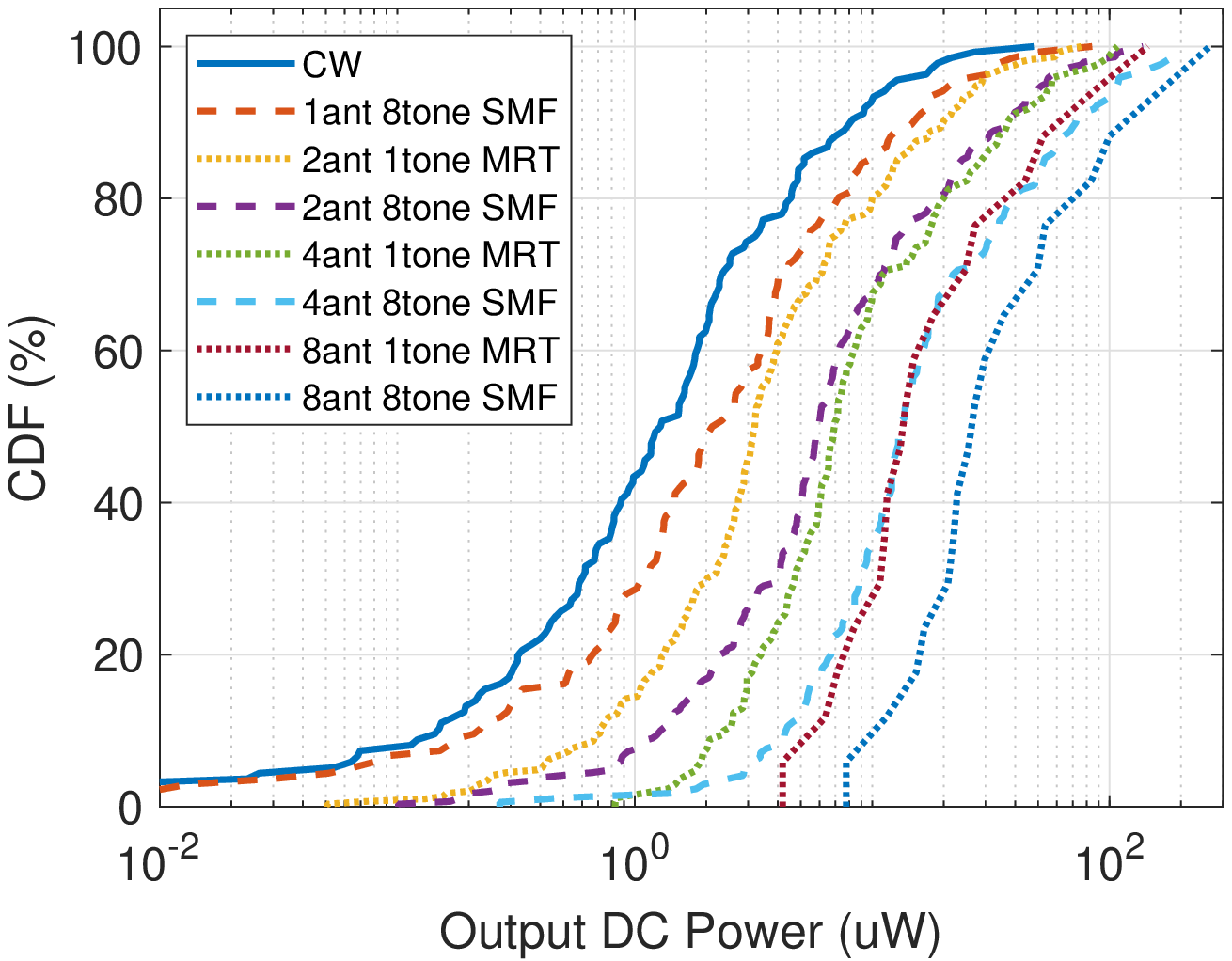}
	\caption{CDF of output DC power measurement results in different distances from 0.6 to 5.4 m (1-, 2-, 4- and 8-antennas, 1- and 8-tones).}
	\label{cdf}
  \end{minipage}
  \hspace*{\fill}
  \begin{minipage}[b]{0.32\linewidth}
	\centering
	\includegraphics[width=0.91\textwidth]{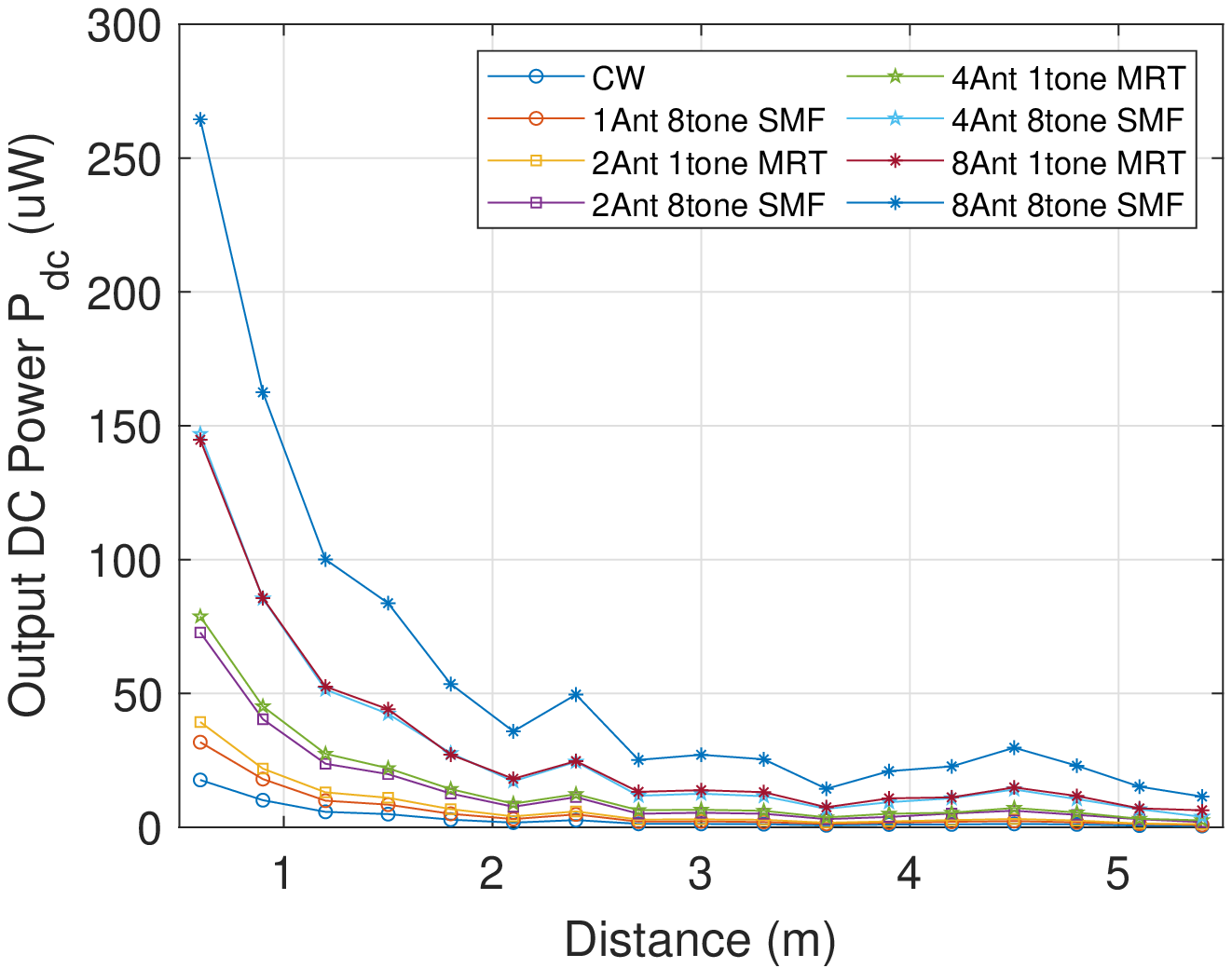}
	\caption{Measurement results of average output DC power versus distances in ranges from 0.6 to 5.4 m (1-, 2-, 4- and 8-antennas, 1- and 8-tones).}
	\label{distance}
  \end{minipage} 
  \hspace*{\fill}
  \begin{minipage}[b]{0.32\linewidth}
	\centering
	\includegraphics[width=0.91\textwidth]{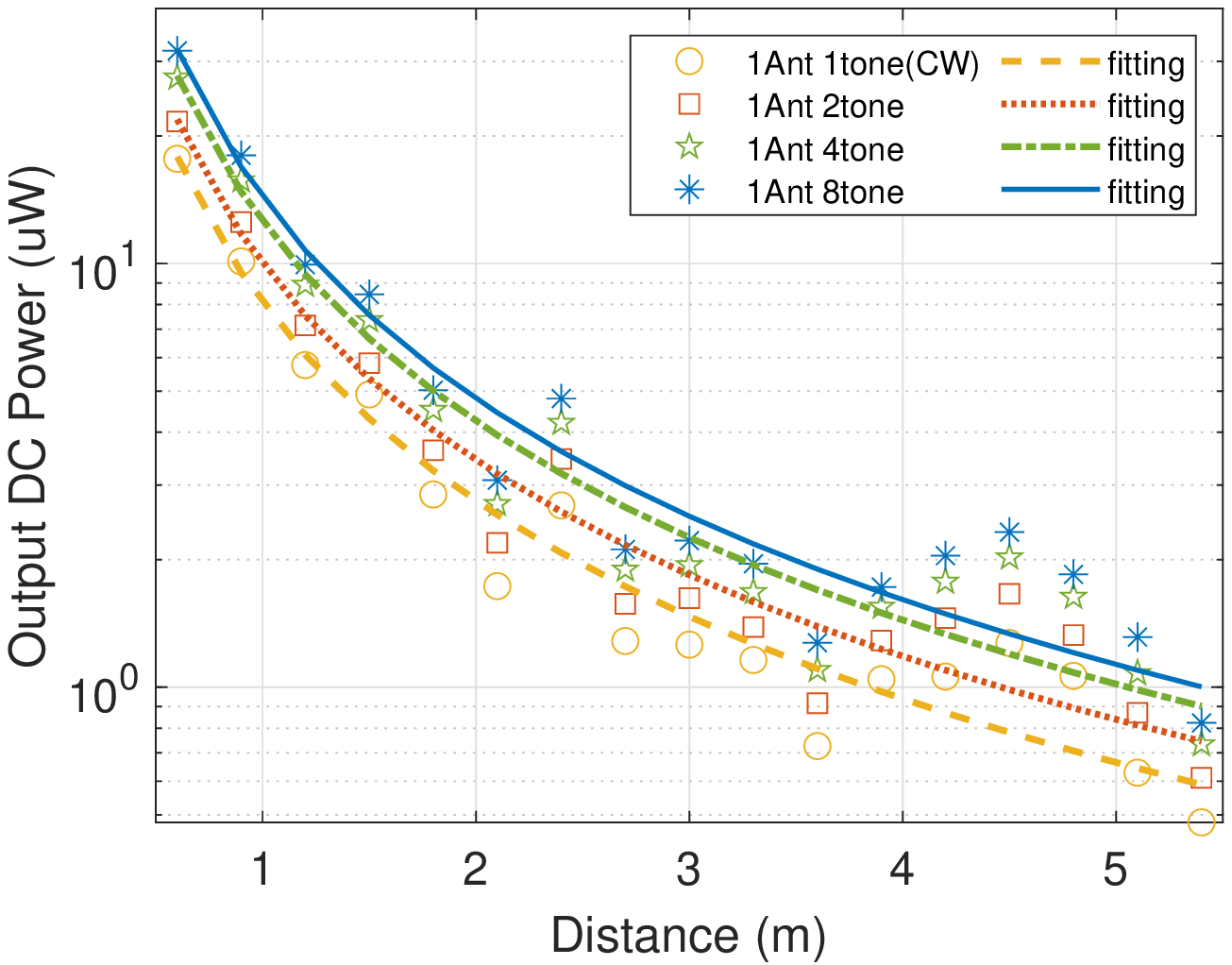}
	\caption{Curve fitting of output DC power ($P_{\mathrm{dc}}$) versus distance ($d$) for SMF signals with 1, 2, 4, 8 tones ($N$=1,2,4,8) and single transmit antenna ($M$=1).}
	\label{fit_tones}
  \end{minipage} 

  \begin{minipage}[b]{0.32\linewidth}
    	\centering
	\includegraphics[width=0.91\textwidth]{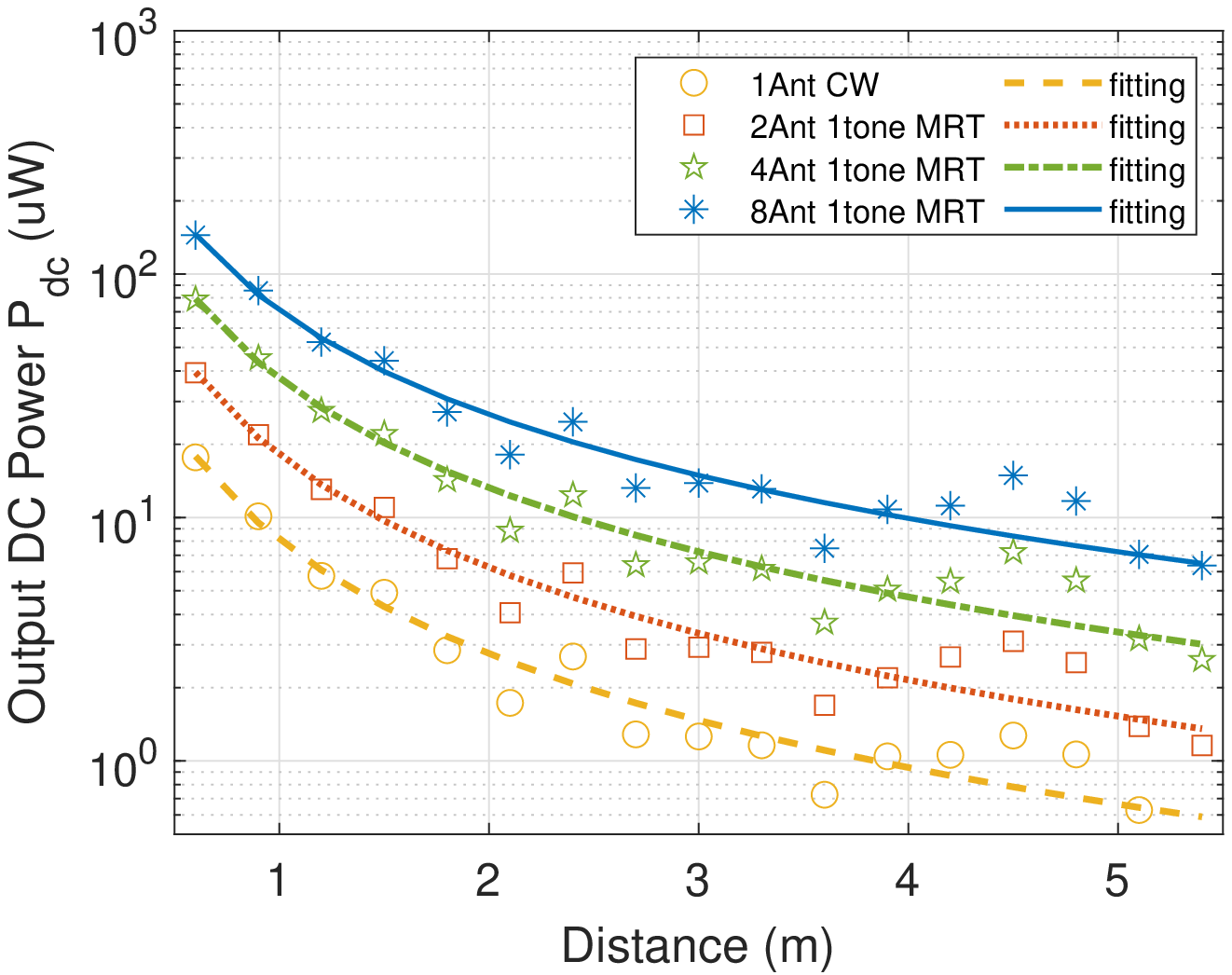}
	\caption{Curve fitting of output DC power ($P_{\mathrm{dc}}$) versus distance ($d$) for MRT beamforming and UP signals with 1-tone ($N$=1) and multiple transmit antennas ($M$=1,2,4,8).}
	\label{fit_antennas}
	\vspace{-2ex}
  \end{minipage}
  \hspace{\fill}
  \begin{minipage}[b]{0.32\linewidth}
    	\centering
	\includegraphics[width=0.91\textwidth]{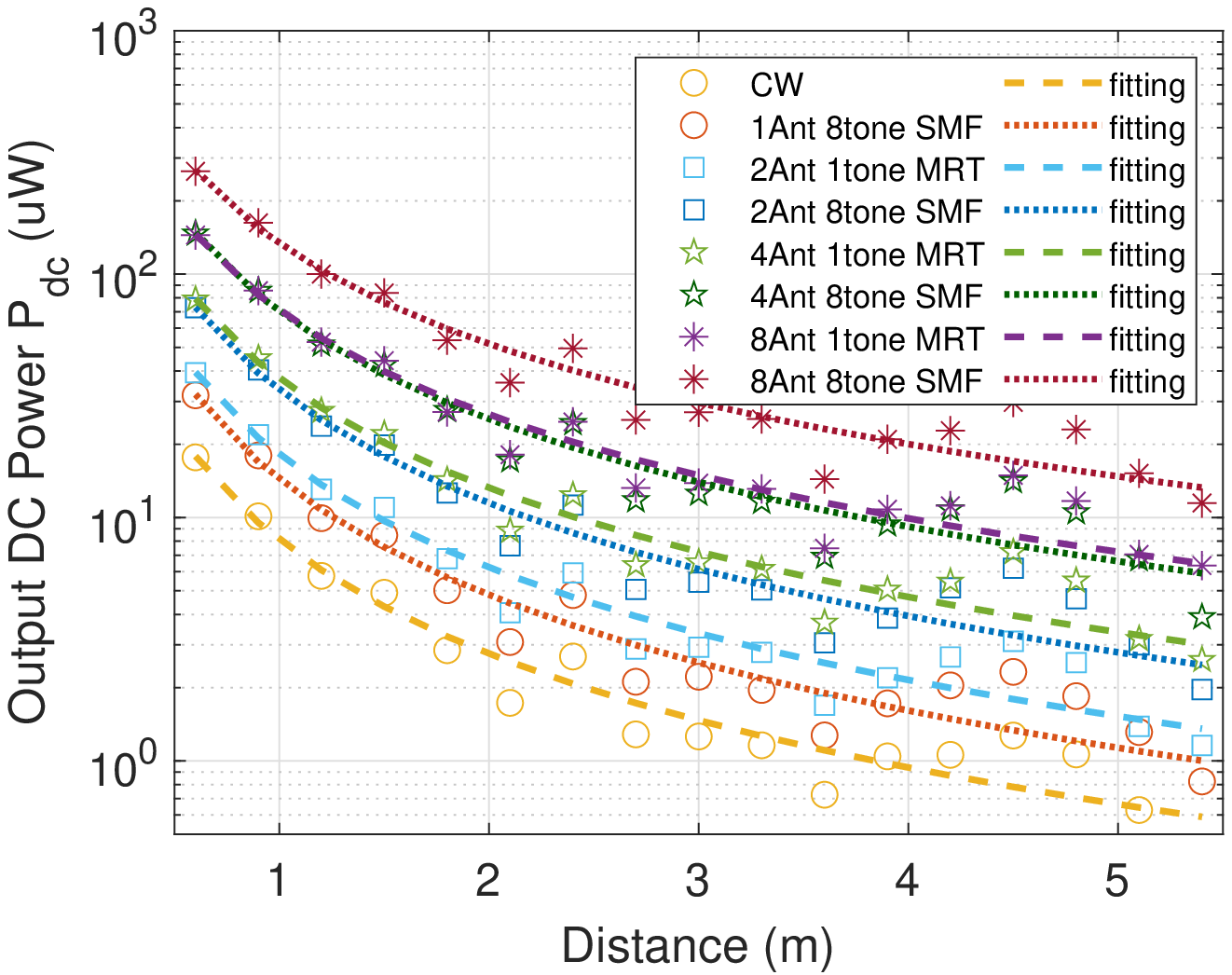}
	\caption{Curve fitting and measurement results of output DC power ($P_{\mathrm{dc}}$) versus distance ($d$) for CW, MRT and SMF signals with $N$=1, 8 and $M$=1,2,4,8.}
	\label{fit_all}
\vspace{-2ex}
  \end{minipage} 
  \hspace{\fill}
  \begin{minipage}[b]{0.32\linewidth}
    	\centering
	\includegraphics[width=0.93\textwidth]{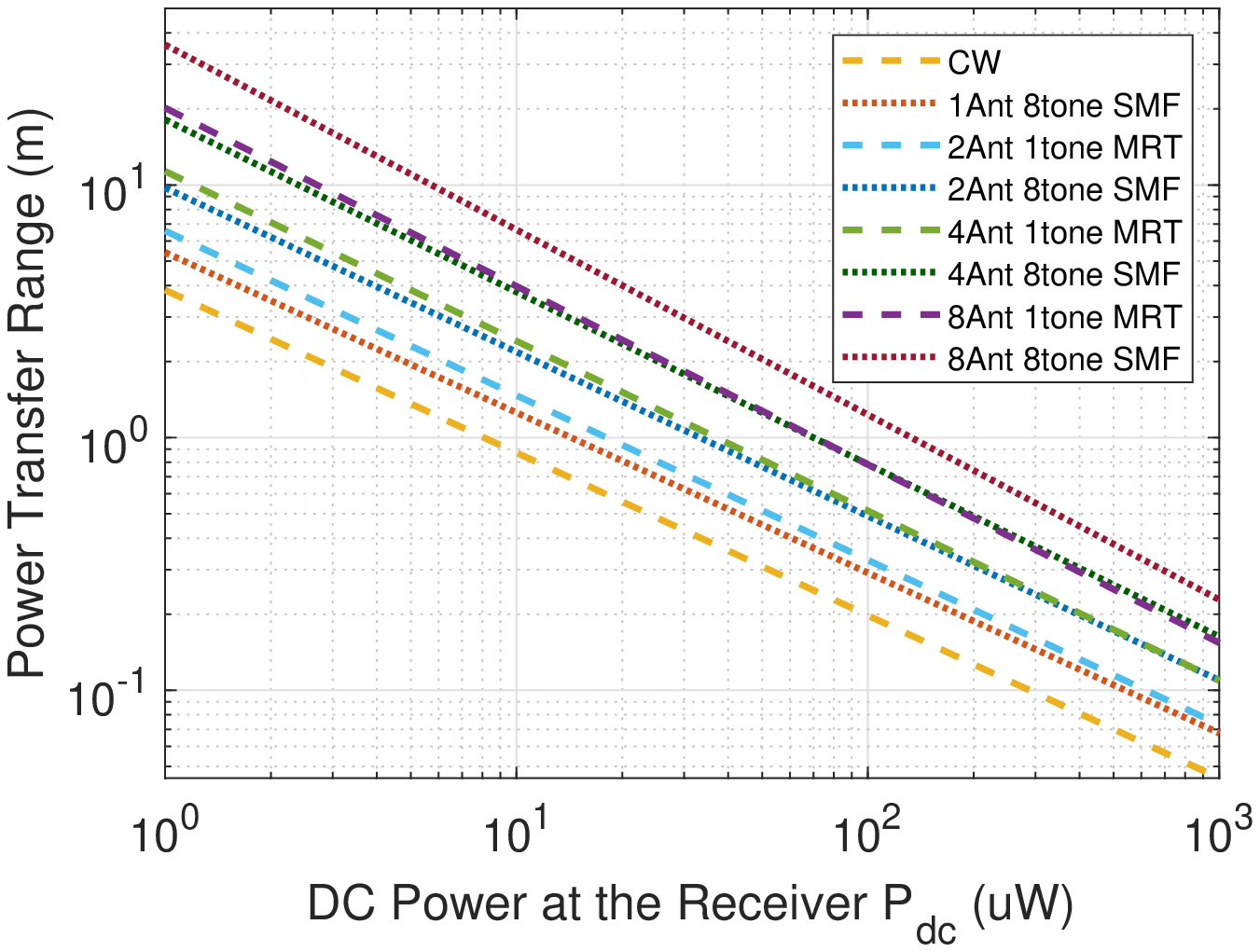}
	\caption{Achievable power transfer range versus required output DC power ($P_{\mathrm{dc}}$) at the receiver based on the fitted model.}
	\label{range}
\vspace{-2ex}
  \end{minipage}

\end{figure*}
%

\vspace{-0.2 cm}
\section{Performance Analysis with Curve Fitting.}  \label{analysis}

In this section, we experiment the WPT in a real-world indoor setting using the beamforming and waveform strategies, collect data and use curve-fitting to draw some conclusions on the effect of beamforming, waveform, and distance on the WPT performance. 
Harvested DC powers at the receiver are measured around one hundred different indoor locations. 
The distance between transmitter and receiver varies from 0.6m to 5.4m. 
We used three different types of WPT signals which are channel non-adaptive CW, single-tone and channel-adaptive multi-antenna MRT beamforming, and channel-adaptive multi-tone and multi-antenna SMF ($\beta{=}3$). 
The number of tones and antennas used in the experiment vary as 1, 2, 4, and 8.
\par
Cumulative distribution functions (CDF) of the measured output DC power levels for the different WPT signals are shown in Fig. \ref{cdf}. 
We only display the 1 and 8 tones results for each antenna setup so as to keep the figures readable.
The CDF graph clearly shows that the measurement results using channel-adaptive signals with more tones and transmit antennas are shifted towards higher output DC power ranges.
When the number of antennas or tones are fixed, using signals with more tones or more antennas, respectively, enhances the output DC power distribution ranges. These observations indicate that WPT performance gains can be obtained by exploiting either the frequency domain, the spatial domain, or both domains jointly. 
Besides, the 8-antenna 1-tone waveform shows a similar performance to that of the 4-antenna 8-tone waveform.
In the same manner, 4-antenna single-tone and 2-antenna 8-tone, and 2-antenna single-tone and 1-antenna 8-tone show similar performance. 
Such behavior demonstrates that one can trade the spatial domain (number of antennas) processing with the frequency domain (number of tones) processing and inversely, and the gains can be accumulated using a joint beamforming and waveform strategy.
\par
Leveraging those measurement data, we model the relationship between the output DC power and the power transfer distance to analyze the benefits of beamforming and waveform strategies and combination thereof to expand the WPT range.
The average measurement results of the harvested DC power versus the distance between the transmitter and the receiver for different signal design schemes, tones and antennas are shown in Fig. \ref{distance}.
All measurement results show an exponentially decreasing trend as the distance increases, which is consistent with common knowledge.
Based on the collected measurement data, we could choose a model that fits the data. 
We fitted the data using a power function as follows
\vspace{-0.2 cm}
\begin{equation}
P_{\mathrm{dc}} = ad^{b}
\end{equation}
where $d$ refers to distances and $a$, $b$ are coefficients to be fitted.

\par
Fig. \ref{fit_tones} shows curve fitting of the output DC power results for SMF signals with a different number of tones and a single antenna transmitter. 
We plot the output DC power ($P_{\mathrm{dc}}$) in the log scale in order to better distinguish the results visually.
The coefficients $a$ and $b$ of the power functions are given as $a{=}8.081$, $b{=}-1.553$ for 1-tone signal, $a{=}9.975$, $b{=}-1.538$ for 2-tone signal, $a{=}12.52$, $b{=}-1.560$ for 4-tone signal, and $a{=}14.32$, $b{=}-1.577$ for 8-tone signal. 
The fitting results show that the exponent $b$ does not vary significantly with the number of tones, and its value is about $-1.5$.
On the other hand, the coefficient $a$ slightly varies with the number of tones.
Approximately, $a$ is increased by 25\% when the number of tones is doubled, and it roughly doubles when $N$ is increased from 1 to 8.
Similar trends are observed from the measurement and fitting results of 2, 4 and 8 antennas cases.
Thus, the expected harvested DC power gain from the frequency domain processing using the waveform strategy is about 25\% when the distance is fixed and the number of frequencies is doubled.
\par
Fig. \ref{fit_antennas} shows the curve fitting of the output DC power results for single-tone MRT signals with different numbers of transmit antennas. 
The coefficients $a$ and $b$ of the power functions are given as $a{=}8.081$, $b{=}-1.553$ for 1-antenna signal, $a{=}18.05$, $b{=}-1.535$ for 2-antenna signal, $a{=}37.07$, $b{=}-1.488$ for 4-antenna signal, and $a{=}70.97$, $b{=}-1.417$ for 8-antenna signal. 
The fitting results for the different numbers of antennas also show no noticeable change in the exponent $b$.
However, the coefficient $a$ shows significant differences between the different numbers of antennas. 
The coefficient $a$ roughly doubles when the number of antennas is doubled.
In other words, 100\% gain in harvested DC power at the receiver can be obtained using a beamforming when the distance is fixed and the number of antennas is doubled. 
\par
The measurement and fitting results confirmed that significant gains at the same distance could be obtained as $N$ and $M$ increase through waveform and beamforming strategies, respectively.
Remarkably, the following results show that the gains are cumulative.
Fig. \ref{fit_all} displays the measured output DC power and their fitting results for multi-tone multi-antenna SMF signals.
It is noted that $P_{\mathrm{dc}}$ can be doubled thanks to the waveform gain when the number of tones is increased from 1 to 8 and to the beamforming gain when the number of transmit antennas is doubled.
In other words, the power that can be harvested from the 8-tone with $M$ antenna signal is the same as the power that can be harvested from the CW signal with $2$$\times$$M$ antennas.
In Fig. \ref{fit_all}, such behavior is well illustrated through the fitting graph.
For instance, the 4-antenna 8-tone curve and the 8-antenna 1-tone curve almost overlap.
These results confirm that the gains from the spatial (beamforming) and frequency (waveform) domains can be jointly accumulated.
The observations of output DC power at the receiver with the same distance are consistent with the scaling laws as derived based on the theoretical non-linear rectifier model.
The scaling laws predicted that the output power is proportional to $M$ in its second-order term and $NM^{2}$ in its fourth-order term and the number of $N$ and $M$ jointly influence the overall performance.  
\par
The observations that the gain in harvested DC power can be improved by increasing $N$ and $M$ with fixed distance signify that the power transfer range is also able to be expanded using multi-tone and multi-antenna when the target power at the receiver is fixed.  
The modeled function can be inverted to express how the target power affects the WPT range and can be written as
\begin{equation} \label{eq_range}
d = 10^{\frac{\mathrm{log}(P_{\mathrm{dc}})-\mathrm{log}(a)}{b}},
\end{equation}
where the coefficient $a$ and $b$ are the same as those obtained from the previous fitting.
The relationship between WPT range and target harvested DC power at the receiver for different signals with different $N$ and $M$ is graphically presented in Fig. \ref{range} based on \eqref{eq_range}.
In line with the previous observation on harvested DC power, this observation shows that the power transfer range can also be improved with increasing $N$ and $M$.
The model shows the range gain is roughly 15\% when the number of tones $N$ is doubled, and approximately 60\% when the number of antennas $M$ is doubled. 
These frequency and spatial domain gains on the power transfer range are cumulative, similarly to the gains observed in the harvested DC power.
So the 8-tone signals with a certain number of antennas show similar range curve as 1-tone with a number of antennas twice as large, as illustrated in Fig. \ref{range}.
For example, the power transfer range can be expanded four times compared to CW signal by signal designs using two different combinations of spatial or frequency domain gains, such as 8-antennas MRT signal or 4-antennas with 8-tone SMF signal. 
It can indeed be seen in Fig. \ref{fit_all} that the output DC power of 1-antenna CW signal at 1m distance is the same as the output DC power of 4-antenna and 8-tone SMF signal and 8-antenna and 1-tone MRT signal at 4m distance.
This example clearly shows that the expansion of the power transfer range using the cumulative gains from the frequency and spatial domains predicted by the fitting result is consistent with the actual measurement result.
\vspace{-0.7 cm}
\section{Conclusions} \label{conclusion}
We implemented a MISO WPT system equipped with up to eight antennas and experimentally evaluated the output DC power performance using channel-adaptive signals at various locations in an indoor laboratory environment.
The measured results were curve-fitted to model the WPT performance according to the distance, and the gains from beamforming and waveform strategies were calculated. 
In addition, it was confirmed that gains in the frequency and spatial domains are cumulative. 
Thanks to the joint beamforming and waveform design and its cumulative gains, it has been verified that the output DC power can be greatly improved. 
It has been shown that the gains from the waveform and beamforming strategies not only improves the output DC power that can be harvested at the same location but also enables power transfer to a longer distance with the same average transmit power.
The relationship between the power transfer range expansion and factors such as number of tones and antennas was simply expressed through a curve-fitting model and confirmed by comparison with the actual output DC power measurement results. The curve-fitting model enables to predict the performance of WPT deployments and can therefore be used to dimension future WPT networks.
Many interesting topics for future research arise from the results of this paper. 
Reducing energy cost for CSI acquisition is an important challenge to enable RF WPT as a practical power source for IoT devices.
It is also worth to implement a large-scale WPT system with multi-tone, multi-antenna, and multi-user. 
\vspace{-0.4 cm}

\ifCLASSOPTIONcaptionsoff
  \newpage
\fi

\bibliographystyle{IEEEtran}
\bibliography{jhlib}

\end{document}